\documentclass[useAMS]{mn2e}
\usepackage[dvips]{graphicx}
\usepackage{longtable}

\def\kms{\mbox{$\rm km~s^{-1}$}}

\def\deg{$^\circ$~}
\def\arcsec{^{\prime\prime}~}

\begin{document}

\title[X-ray Bright Nucleus in LSB galaxy UGC~6614]
{An X-ray Bright Nucleus in the Low Surface Brightness Galaxy UGC~6614}

\author[Naik et al.]{Sachindra~Naik$^{1}$, M.~Das$^{2,4}$, C.~Jain$^{2,3}$ and B.~Paul$^{2}$\\
1.~Physical Research Laboratory, Navrangpura, Ahmedabad 300 009, India\\
2.~Raman Research Institute, Sadashivanagar, Bangalore 560080, India\\
3.~Department of Physics and Astrophysics, University of Delhi, Delhi 110007, India\\
4.~Birla Institute of Technology and Science - Pilani, Hyderabad Campus, Jawahar Nagar, Shameerpet
Mandal, Hyderabad, 500078, India\\
(E-mail~:~snaik@prl.res.in)}

\date{Accepted for publication in MNRAS}

\maketitle


\begin{abstract}
We report a study of the X-ray emission from the nuclear region of the 
low surface brightness (LSB) galaxy UGC~6614. Very little is known about 
the central objects in LSB galaxies especially their X-ray properties and 
X-ray spectra. In this study we have used \emph{XMM-Newton} archival data
to study the characteristics of the X-ray spectrum and the X-ray flux
variability of the AGN in the LSB galaxy UGC~6614. The nucleus of UGC~6614 
is very bright in X-ray emission with an absorption corrected 0.2-10.0 keV
luminosity of $\sim$1.1$\times$ 10$^{42}$ erg s$^{-1}$. The X-ray spectrum 
is found to be power-law type with a moderate column density. A short time 
scale of intensity variation and large X-ray flux is indicative of the 
presence of a black hole at the centre of this galaxy. Using the method 
of excess variance, we have determined the black hole mass to be 
$\sim0.12\times10^{6}~M_{\odot}$. The X-ray spectral properties are similar 
to that of the Seyfert I type AGNs. Our study thus demonstrates that although
LSB galaxies are poor in star formation, they may harbour AGNs with X-ray
properties comparable to that seen in more luminous spiral galaxies.

\end{abstract}

\section{INTRODUCTION}

Most of the X-ray bright AGN in our nearby universe ($z~<~1$) are either 
bulge dominated early type galaxies or disk galaxies that have recently
undergone minor mergers or bar instabilities (Georgakakis et al. 2009; 
Gabor et al. 2009; Pierce et al. 2007). The latter processes are accompanied 
by vigorous star formation and gas infall towards the nucleus. Hence the 
host galaxies are optically bright and their nuclei often show strong AGN
activity (Ho et al. 2003). Late type spiral galaxies are however not as 
bright as early type galaxies. They are generally poor in star formation 
and if AGN activity is present, it is usually weak (Ho 2008). The nuclear
activity of such late type systems are not as well studied or understood 
as those found in bright galaxies.

In this paper we present a study of the X-ray spectrum of an extreme type 
of late type spiral galaxy, a Low Surface Brightness (LSB) galaxy. These
galaxies are poorly evolved, halo dominated systems and have diffuse, 
optically dim disks (Impey \& Bothun 1997). AGN have been detected in 
their nuclei at optical wavelengths (Sprayberry et al. 1995; Impey, 
Burkholder \& Sprayberry 2001), at radio frequencies (Das et al. 2006; 
2007) and in X-ray emission (Das et al. 2009a). AGN in LSB galaxies are 
usually associated with bulge dominated LSB galaxies (Schombert 1998). 

The X-ray properties of LSB galaxies are relatively unexplored. A recent 
Chandra study of eight giant LSB galaxies detected X-ray emission from the 
AGN in two galaxies (Das et al. 2009a) as well as diffuse emission 
associated with the bulge in four of the eight sources. To date only 
two LSB galaxies have been observed by \emph{XMM-Newton}, UGC~6614 and 
F568-6 (Malin~2). Of the two, UGC~6614 has an X-ray bright core. In F568-6, 
only a very faint X-ray source was detected in the XMM-Newton image; the
faintness is probably due to the larger distance of the galaxy 
($D_{Mpc}\sim200$) compared to that of UGC~6614 ($D_{Mpc}\sim93$, where 
$D_{Mpc}$ is the luminosity distance of the galaxy). 

This paper presents the first study of the nuclear X-ray spectrum of an 
LSB galaxy. X-ray imaging studies of LSB galaxies have been done earlier 
(Das et al. 2009), but X-ray spectroscopic studies have not been done before.
Using \emph{XMM-Newton} archival data we present a study of the nuclear 
X-ray spectrum of UGC~6614 which is a prototypical giant LSB galaxy. The 
galaxy is close to face on and has a prominent bulge surrounded by a ring 
like feature (Rahman et al. 2007; Hinz et al. 2007). Its optical spectrum 
shows a broad $H\alpha$ line typical of a Seyfert 1 nucleus (Schombert 1998). 
It has a bright nucleus at millimeter and centimeter radio wavelengths (Das 
et al. 2006; 2009b). The optical R band image of the galaxy with the contours 
of X-ray emission overlaid is shown in Figure~1 and the galaxy parameters 
are listed in Table~1. In the following sections we describe our analysis 
of the spectral and temporal properties of the X-ray spectrum of UGC~6614. 
We use the latter to estimate the black hole mass associated with the AGN 
and in the last section we discuss the implications of our results.

\vspace{5mm}
\begin{figure}
\includegraphics[width=80mm,height=80mm,angle=0]{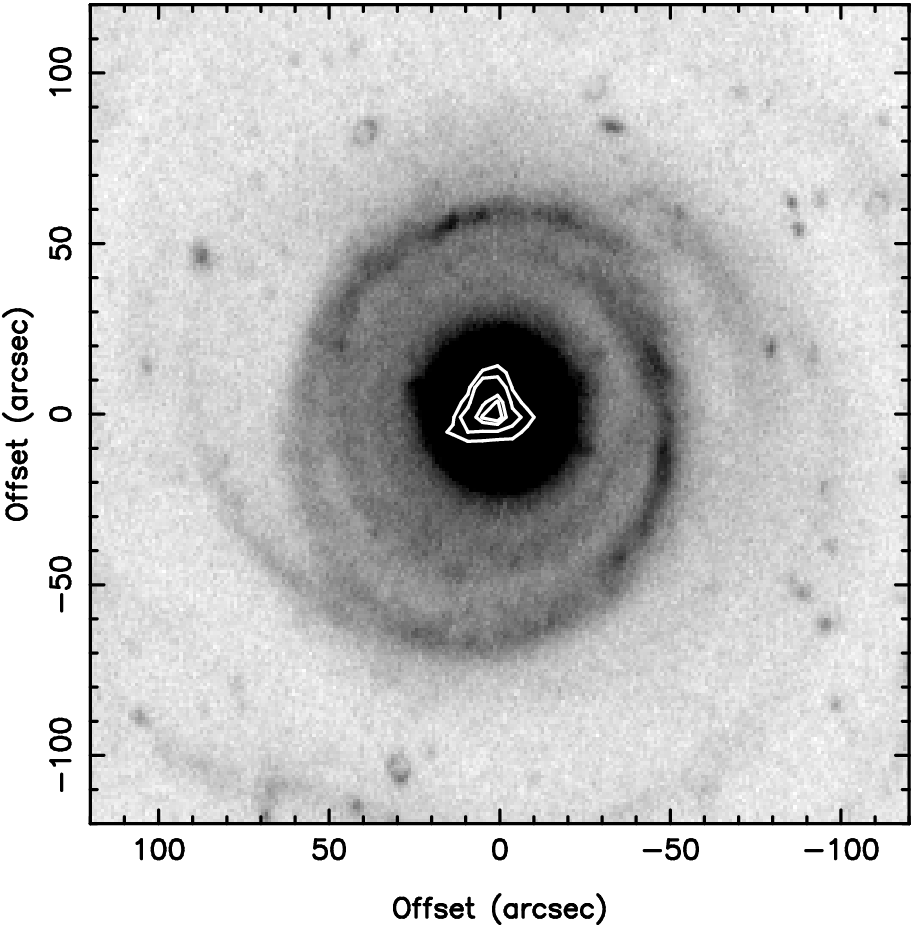}
\caption{The figure shows the optical R band image of the galaxy with the 
contours of X-ray emission overlaid. }
\end{figure}   

\begin{table}
\caption{Parameters of UGC~6614}
\begin{tabular}{ll}
\hline
Parameter	&	Value\\
\hline
Galaxy Type 	& (R)SA(r)a (NED) \\
Galaxy Position (RA, DEC) & 11:39:14.9, +17:08:37 (2MASS) \\
Velocity, Redshift & 6352~\kms, 0.0212 (NED) \\
Distance & 93.3~Mpc (NED) \\
Linear Distance Scale & 0.45~kpc/arcsec \\
Disk Inclination & 29.9\deg (Hyperleda) \\
Disk Optical Size ($D_{25}$) & 99.6$\arcsec$ (RC3) \\
\hline
\hline
\end{tabular}
\label{Table 1}
\end{table}

\section{Data analysis and results}

UGC~6614 was observed with the \emph{XMM-Newton} on 13 June 2002 for an 
exposure of $\sim$13.4 ksec. The raw events were processed using the 
Science Analysis System (SAS) package v7.0.0. Light curves in the energy 
range greater than 10 keV were extracted from PN and MOS event data 
(pattern zero only) to identify intervals of flaring particle background. 
A flare type feature was seen in the hard X-ray light curves ($>$ 10 keV) 
of PN and MOS event data which was subsequently excluded from further 
analysis. Light curves at different energy bands (0.2-2.0 keV, 2.0-4.0 
keV, 4.0-10.0 keV, and 0.2-10.0 keV) were extracted from PN and MOS event 
data by using the SAS task {\it xmmselect} and selecting circular regions 
of radii 30 arcsec and 60 arcsec centered at the source position respectively.
The energy spectra were extracted by selecting the corresponding circular regions around the source. The background light curves and spectra were extracted from the nearby source free regions by selecting multiple circular regions of radii in 30-60 arcsec. The corresponding EPIC responses and 
effective area files were generated by using the SAS tasks {\it rmfgen} and 
{\it arfgen}, respectively. 

\begin{figure}
\includegraphics[width=80mm,height=85mm,angle=-90]{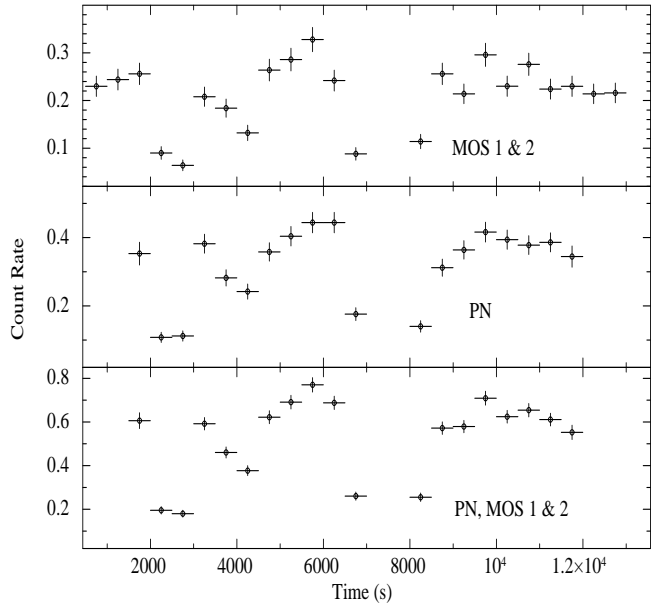}
\caption{The figure shows the 0.2-10 keV light curves of UGC~6614 using 
data obtained from MOS-1 and MOS-2 (upper panel), PN (middle panel) and 
MOS-1, MOS-2 and PN detectors (bottom panel). The light curves are plotted
for 500 s bin time. The missing data points in $\sim$7-8 ksec interval is
due to the presence of a flaring event like feature in the $\geq$10 keV
light curve obtained only from the single events.}
\end{figure}

\begin{figure}
\includegraphics[height=85mm, width=70mm, angle=-90]{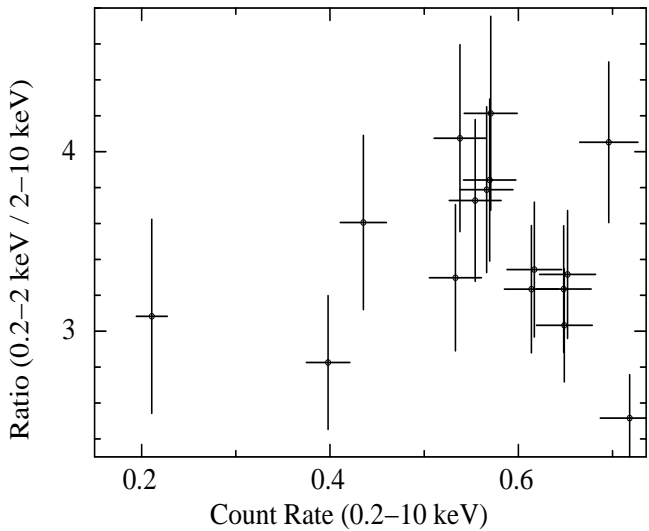}
\caption{The figure shows the ratio between the source count rates in 
0.2-2 keV and 2-10 keV range as a function of 0.2-10 keV count rate. The 
data from MOS-1, MOS-2 and PN detectors are added together.}
\end{figure}

\subsection{Timing analysis}

Light curves at different energy bands, extracted from MOS-1, MOS-2, and 
PN event data as described above, were first combined to increase the 
signal to noise ratio. To investigate whether the source exhibits 
significant variations or not, we used light curves in 0.2$-$2.0 keV, 
2.0$-$4.0 keV and 4.0$-$10.0 keV energy ranges with a bin size of 500 s, 
and fitted with a constant. Given the count rate of the source, the number 
of photons in each bin was always sufficiently high for the use of the 
$\chi^2$ test to investigate whether the source flux remained constant 
or not.  In all the cases, the value of the reduced $\chi^{2}$ was found 
to be greater than 15, thus rejecting the constant hypothesis with a high
significance. This implies that the observed variations are intrinsic to 
the source. However, it should be remarked that at smaller timescales, due 
to poor statistics, it is not possible to establish the presence or absence 
of variability using the $\chi^{2}$ test. In Figure~2, light curves in 
0.2-10 keV energy range, obtained from MOS-1, MOS-2, and PN detectors are 
shown for 500 s time bin.  The added light curve from MOS-1 and MOS-2 is 
shown in the top panel, whereas the light curve obtained from PN detector 
is shown in the middle panel and the added light curve from all three 
detectors is shown in the bottom panel. The variability in the light curves 
in all three panels can be seen clearly. 

In order to check the presence or absence of any flux related spectral 
variations in UGC~6614, we plotted the ratio between the source count 
rates in 0.2-2 keV and 2-10 keV ranges as a function of 0.2-10 keV count
rate. The resultant plot is shown in Figure~3. The light curves from all 
three detectors were added together in corresponding energy ranges to 
improve the signal-to-noise level. From the figure, however, it is difficult
to draw any concluson on the spectral steepening with increasing flux, as
seen in case of Seyfert galaxies.

\subsection{Excess Variance and Black Hole Mass Estimation}

The black hole mass (M$_{BH}$) is a key parameter in the study of AGNs. 
It can be determined by several methods of which one of the most reliable
is the reverberation mapping technique (Peterson et al. 2004). It can also 
be determined from the X-ray variability of AGN emissionn using the break 
in the power spectrum (Papadakis 2004; McHardy et al. 2005) or the excess
variance (Nikolajuk, Papadakis $\&$ Czerny 2004). In this section we use 
the method of excess variance to determine the mass of the black hole. This
method is useful to estimate M$_{BH}$ in AGNs with short data coverage.

Excess variance is defined as the amplitude of variability in the light 
curve in excess of that expected from statistical fluctuations in the 
background level. This excess of the mean-squared normalized light curve 
is defined as (Nandra et al. 1997; Turner et al. 1999):
\begin{equation}
\sigma^{2} = \frac{1}{N \mu^{2}} \sum_{i=1}^{N} \left[ (X_{i} - \mu)^{2} - \sigma_{mean}^{2}\right]
\end{equation}
where, X$_{i}$ is the count rate of N points in the light curve with an 
unweighted arithmetic mean of $\mu$ and $\sigma_{mean}^{2}$ is the mean 
square error in the dataset. The square root of the normalised excess 
variance (F$_{amp}$), indicates the root mean square variability amplitude.   

The excess variance of the light curve between frequencies $\nu_{1}$ and 
$\nu_{2}$ can be calculated from the integral of the power density spectra. 
It is given by (Nikolajuk et al. 2004):
\begin{equation}
\sigma^{2} = \int_{\nu{1}}^{\nu_{2}} P(\nu) d\nu
\end{equation}

The determination of M$_{BH}$ is based on the assumption that the high 
frequency break in the power spectrum ($\nu_{bf}$) is inversely related 
to the black hole mass (M$_{BH}$), and the value of the product of power 
(P($\nu$)) and frequency ($\nu$) at the high frequency break is constant
for all the sources and is independent of their mass. Therefore, as 
mentioned in Nikolajuk et al. (2004),
\begin{eqnarray}
\nu_{bf} = C_{1}/M_{BH}\\
P(\nu_{bf}) \nu_{bf} = C_{2}
\end{eqnarray}
where, C$_{1}$ and C$_{1}$ are constant. Thus, the mass of the black hole 
can be estimated if the excess variance of variability is known in a given
frequency band, above the break frequency.  If T is the duration of the 
light curve and $\Delta$t is the binsize of the light curve, then the mass 
of the black hole is given by:
\begin{equation}
M_{BH} = C \frac{T-2\Delta t} {\sigma^{2}}
\end{equation}
where, C is a constant which is determined by applying this method to a 
standard source. Using Cyg X-1 as a reference object of mass 20 M$_{\odot}$
(Ziolkowski 2005), Nikolajuk et al. (2006) determined the value of C to be 
1.92$\pm$0.5 M$_{\odot}$s$^{-1}$. Nikolajuk et al. (2006), though,
used light curves in 2-10 keV band for their calculations, the use of light
curves in different energy bands in the present case do not affect the
final result as $\sigma^{2}$ is found to be similar at all energy bands.

Using the above method of excess variance, we have determined the normalized
excess variance in light curves binned with 500 s, in different energy bands.
Table~2 gives the estimated values of the excess variance ($\sigma^{2}$), the
variability amplitude (F$_{amp}$) and the mass of the black hole (M$_{BH}$) 
in each case. The numbers in the bracket are the errors calculated using the 
F$_{amp}$ statistics described in Vaughan et al. (2003). 

\begin{table}
\caption{X-ray variability parameters for UGC~6614}
\begin{tabular}{l c c c}
\hline
Energy  	& $\sigma^{2}$ 	& F$_{amp}$  	& M$_{BH}$\\
(keV)		&		&		& $\times$ 10$^{6}$ M$_{\odot}$\\
\hline
0.2$-$2.0 	& 0.154(0.011)	& 0.392(0.014)	& 0.11(3)\\
2.0$-$4.0 	& 0.137(0.025)	& 0.369(0.034)	& 0.13(6)\\
4.0$-$10.0 	& 0.142(0.034)	& 0.376(0.046)	& 0.12(6)\\
\hline
\end{tabular}
\end{table}

\subsection{Spectral Analysis}

Source and background spectra, response files and the effective area 
files for each of the three detectors were generated as described above. 
The source spectra were binned to a minimum of 20 counts per bin to 
improve the statistics and ensure the validity of the $\chi^2$ 
test. The spectral fitting was performed with the 
XSPEC package (v11). After appropriate background subtraction, 
simultaneous spectral fitting was done using the MOS1, MOS2, and PN 
spectra. All the spectral parameters other than the relative normalization, 
were tied toegther for all three detectors. We fitted the 0.1-10.0 keV 
spectra with a power-law continuum component modified with interstellar
absorption. The model was found to be statistically acceptable with
a reduced $\chi^2$ of 1.1 for 225 degrees of freedom. The value of 
the $N_H$ was found to be $\sim$2 $\times$ 10$^{20}$ atoms cm$^{-2}$, 
which is similar to that of the Galactic value in the source direction.
Therefore, we fixed the value of $N_H$ at the Galactic value (1.93 $\times$
10$^{20}$ atoms cm$^{-2}$) and estimated the best-fit parameters. The 
energy spectrum of the source along with the residuals to the best-fit 
model is shown in Figure~4. The spectral parameters of the best-fit model 
obtained from the simultaneous spectral fitting are given in Table 3.

\begin{figure}
\includegraphics[width=70mm,height=85mm,angle=-90]{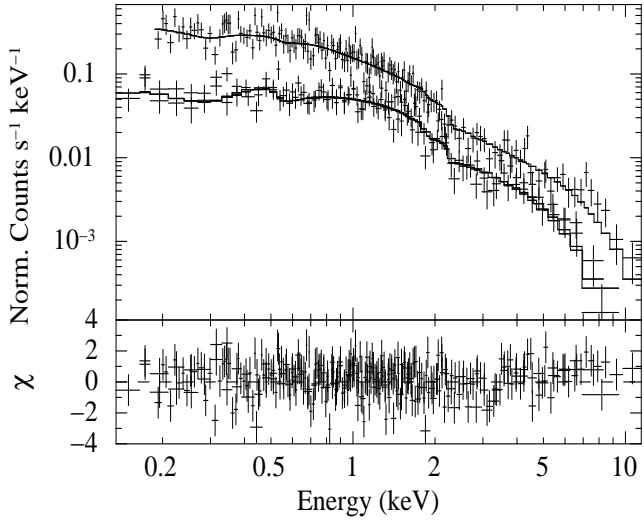}
\caption{Energy spectrum of UGC~6614 obtained with the MOS-1, MOS-2, 
and PN detectors of the \emph{XMM-Newton}, along with the best-fit model
comprising a power-law continuum component and interstellar absorption. 
The bottom panel shows the contributions of the residuals to the $\chi^2$ 
for each energy bin.}
\end{figure}

\begin{table}
\caption{Spectral parameters for UGC~6614}
\noindent
\begin{tabular}{ll}
\hline
\hline
Power-law index ($\Gamma$)           &1.81$\pm$0.02\\
Absorption uncorrected Source flux$^a$  &1.03$\pm$0.02 \\
Absorption corrected source flux$^a$ &1.13$\pm$0.02 \\
Reduced $\chi^2$                     &1.1 (297 dof)\\
\hline
\multicolumn{2}{l}{
Note: Errors are defined in 1$\sigma$ confidence limit.}\\
\multicolumn{2}{p{6cm}}{
$^a$ : Flux (in 10$^{-12}$ ergs cm$^{-2}$ s$^{-1}$) is estimated in
0.2--10~keV energy range.}\\
\hline
\hline
\end{tabular}
\label{spec_par}
\end{table}

\section {Discussion}

The main result of this paper is that although UGC~6614 is a LSB galaxy, 
it hosts an AGN whose X-ray charactersitics are similar to those found in 
more evolved star forming galaxies and Seyfert galaxies. In the following
paragraphs we discuss the implications of our results.

\noindent
{\bf (i) An X-ray bright AGN in a LSB galaxy~:~}\\
The nucleus of UGC~6614 has an X-ray luminosity of $L_{X}\sim1.1\times10^{42}~ergs^{-1}$. This is similar to the X-ray emission observed from another 
giant LSB galaxy, UGC~2936 ($L_{X}\sim1.8\times10^{42}~ergs^{-1}$; Das et 
al. 2009a). The flux is also comparable to the nuclear X-ray flux of nearby disk
galaxies (Georgakakis et al. 2009). Further, like most of the Seyfert galaxies, UGC~6614 shows significant X-ray variability on short timescales (Turner et al. 1999). This strongly indicates the presence of an active nucleus in UGC ~6614. 
As mentioned earlier in Section~1, the nucleus of UGC~6614 is also bright 
at radio and optical frequencies. The radio spectrum is flat between 1.4~GHz 
to 110~GHz and the emission is compact in morphology (Das et al. 2006) and 
only at lower radio frequencies (610~MHz) does it appear extended as radio 
lobes or jets (Das et al. 2009b). Thus, although UGC~6614 is a low luminosity
galaxy whose evolution is slowed down by the presence of a dominant dark halo,
its AGN characteristics are similar to more evolved, star forming galaxies.

\noindent
{\bf (ii)~X-ray Variability and BH mass~:~}\\
The X-ray emission shows variability at short time scales. Intraday X-ray
variability has been observed in most quasars and AGNs (e.g. Markowitz \& 
Edelson 2001; Uttley, Mc Hardy \& Papadakis 2002), while some AGNs also show 
variability down to time scale of minutes (e.g. MCG-6-30-15: Miniutti et al. 2007). The short variability time scale has strong implications; it restricts the size of the emission region and is also related to efficiency of the central engine. The large amplitude, short time scale X-ray variability of UGC~6614, indicates an accreting black hole as the source of nuclear energy in this galaxy. If the X-ray variability mechanism in UGC~6614 and nearby Seyfert galaxies is the same, then, using the method of Nikolajuk et al. (2004), we estimated that the mass of the central black hole is $M_{BH}\sim0.12\times$ 10$^{6} $M$_{\odot}$. This is similar to the black hole mass determined in several low luminosity Seyfert-like active galaxies, such as, the late type spiral galaxy NGC~4395 (Vaughan et al. 2005), the dwarf elliptical POX 52 (Barth et al. 2004) and the dwarf Seyfert 1 SDSS J160531.84+174826.1 (Dong et al. 2007). However, an approximate estimate of $M_{BH}$ the from the H$\alpha$ line luminosity and linewidth relation (Greene \& Ho 2007) gives a value of 
$M_{BH}~=~1.06\times$ 10$^{7} $M$_{\odot}$ (Das et al. 2009a), which is much higher than that derived from the X-ray variability. Both methods are approximate and have large dispersions in the black hole mass. But the H$\alpha$ observations are very poor in resolution (Sprayberry et al. 1995) and can be considerably improved. In fact the larger $M_{BH}$ value is only an upper limit for the mass. This is supported by recent H$\alpha$ spectroscopic observations of the nucleus of this galaxy (Ramya et al. 2010, in preparation). The Eddington luminosity correpsonding to $M_{BH}\sim0.12\times$ 10$^{6} $M$_{\odot}$ is
$L_{Edd}~=~1.7\times10^{43}~ergs~s^{-1}$. The X-ray luminosity is already 10\% of the Eddington limit for this black hole mass. Vasudevan \& Fabian (2009) reported a range of 15 to 150 for the bolometric correction in AGNs depending on the Eddington ratio. Thus the mass of the black hole in UGC~6614 is quite likely to be more than the nominal value determined from excess variance. In absence of a knowledge about the SED of this object, we cannot obtain a more definite estimate for the black hole mass.  

\noindent
{\bf (iii)~Implications for galaxy evolution~:~}\\
Although the existence of AGN and black holes (BHs) in giant LSB galaxies 
has been known for some time, their properties are not well understood. Some
studies suggest that their black hole mass and central velocity dispersion 
($M-\sigma$) relation is different from ellipticals and high surface 
brightness galaxies (Pizzella et al. 2005). They also appear to lie off 
the radio-X-ray fundamental plane (Das et al. 2009). However, more detailed
studies are required to understand the nuclear properties of these galaxies. 
As mentioned earlier giant LSB galaxies have optically dim disks but prominent
bulges which often host relatively bright AGN. This suggests that the bulge and
AGN evolved independently of the halo dominated disks. This decoupled 
disk-nuclear evolution is unlike what we see in nearby star forming
spirals where the bulge-AGN evolution is related to the secular evolution 
of the galaxy as a whole (Kormendy \& Richstone 1995). This type of 
bulge-AGN evolution does not fit into our present picture of galaxy evolution.
To understand it requires a deeper study of the AGN characteristics and star formation history of UGC~6614. 

\section{ACKNOWLEDGMENTS}
The authors thank the referee for his valuable suggestions that helped 
us to improve the content of this paper. We would like to thank Alice Quillen
for providing the R band image of UGC~6614. This work is based on observations
obtained with XMM-Newton, an ESA science mission with instruments and 
contributions directly funded by ESA Member States and the USA (NASA). We 
have also used the NASA/IPAC Infrared Science Archive, which is operated by 
the Jet Propulsion Laboratory, California Institute of Technology, under
contract with the National Aeronautics and Space Administration. The research
work at Physical Research Laboratory is funded by the Department of Space,
Government of India.


\begin{thebibliography}{}
\bibitem[\protect\citeauthoryear{Barth}{2004}]{Barth04}
Barth, A. J., Ho, L. C., Rutledge, R. E., Sargent, W. L. W. 2004, ApJ, 607, 90
\bibitem[\protect\citeauthoryear{Das et al.}{2006}]{Das06}
Das, M., O'Neil, K., Vogel, S. N., McGaugh, S. S., 2006, ApJ, 651, 853
\bibitem[\protect\citeauthoryear{Das et al.}{2007}]{Das07}
Das, M.; Kantharia, N.; Ramya, S.; Prabhu, T. P.; McGaugh, S. S.; Vogel, S. N. 2007, MNRAS, 379, 11
\bibitem[\protect\citeauthoryear{Das et al.}{2009}]{Das09}
Das, M.; Reynolds, C. S.; Vogel, S. N.; McGaugh, S. S.; Kantharia, N. G. 2009a, ApJ, 693, 1300
\bibitem[\protect\citeauthoryear{Das et al.}{2009a}]{Das09a}
Das, M.; Kantharia, N. G.; Vogel, S. N.; McGaugh, S. S. 2009b, in Low Frequency Radio Universe, 
ASPC conference series, 407, 208.
\bibitem[\protect\citeauthoryear{Dong et al.}{2007}]{Dong07}
Dong, X., Wang, T., Yuan, W., Shan, H., Zhou, H., Fan, L., Dou, L., Wang, H., Wang, J., Lu, H. 2007, ApJ, 657, 700
\bibitem[\protect\citeauthoryear{Edelson}{1996}]{Edelson96}
Edelson, R. A.; Alexander, T.; Crenshaw, D. M.; Kaspi, S.; Malkan, M. A.; Peterson, B. M.; Warwick, R. S.; Clavel, J.; Filippenko, A. V.; Horne, K.; and 81 coauthors 1996, ApJ, 470, 364
\bibitem[\protect\citeauthoryear{Gabor09}{2009}]{Gabor09}
Gabor, J. M.; Impey, C. D.; Jahnke, K.; Simmons, B. D.; Trump, J. R.; Koekemoer, A. M.; Brusa, M.; Cappelluti, N.; Schinnerer, E.; Smolčić, V. (and 7 coauthors) 2009, ApJ, 691, 705 
\bibitem[\protect\citeauthoryear{Georgakakis09}{2009}]{Georgakakis09}
Georgakakis, A.; Coil, A. L.; Laird, E. S.; Griffith, R. L.; Nandra, K.; Lotz, J. M.; Pierce, C. M.; Cooper, M. C.; Newman, J. A.; Koekemoer, A. M.  2009, MNRAS, in press (arXiv0904.3747)
\bibitem[\protect\citeauthoryear{GreeneHo09}{2007}]{GreeneHo07}
Greene, J. E.; Ho, L. C. 2007, ApJ, 670, 92
\bibitem[\protect\citeauthoryear{Hinz07}{2007}]{Hinz}
Hinz, J. L., Rieke, G. H., Willmer, C. N. A., Misselt, K., Engelbracht, C. W., Blaylock, M., Pickering, T. E., 2007, ApJ, 663, 895
\bibitem[\protect\citeauthoryear{Ho}{2003}]{Ho03}
Ho, Luis C.; Filippenko, Alexei V.; Sargent, Wallace L. W. 2003, ApJ, 583, 159
\bibitem[\protect\citeauthoryear{Ho}{2008}]{Ho08}
Ho, Luis C. 2008, ARA\&A, 46, 475
\bibitem[\protect\citeauthoryear{ImpeyBothun}{1997}]{ImpeyBothun97} 
Impey, C.; Bothun, G. 1997, ARA\&A, 35, 267
\bibitem[\protect\citeauthoryear{Impey et al.}{2001}]{Impey2001}
Impey, C., Burkholder, V., Sprayberry, D. 2001, AJ, 122, 2341
\bibitem[\protect\citeauthoryear{Kormendy}{1995}]{KormendyRichstone}
Kormendy, John; Richstone, Douglas 1995, ARA\&A, 33, 581
\bibitem[\protect\citeauthoryear{Markowitz01}{2001}]{MarkowitzEdelson}
Markowitz, A.; Edelson, R. 2001, ApJ, 547, 684
\bibitem[\protect\citeauthoryear{Miniutti09}{2007}]{Miniutti07}
Miniutti, Giovanni; Fabian, Andrew C.; Anabuki, Naohisa; Crummy, Jamie; Fukazawa, Yasushi; Gallo, Luigi; Haba, Yoshito; Hayashida, Kiyoshi; Holt, Steve; Kunieda, Hideyo; and 14 coauthors  2007, PASJ, 59, 315
\bibitem[\protect\citeauthoryear{McHardy et al.}{2005}]{McHardy05}
McHardy, I. M., Gunn, K. F., Uttley, P., Goad, M. R., 2005, MNRAS, 359, 1469
\bibitem[\protect\citeauthoryear{Nandra et al.}{1997}]{Nandra97}
Nandra, K., George, I. M., Mushotzky, R. F., Turner, T. J., Yaqoob, T., 1997, ApJ, 476, 70
\bibitem[\protect\citeauthoryear{Nikolajuk et al.}{2004}]{Nikolajuk04}
Nikolajuk, M., Papadakis, I. E., Czerny, B., 2004, MNRAS, 350, L26
\bibitem[\protect\citeauthoryear{Nikolajuk et al.}{2006}]{Nikolajuk06}
Nikołajuk, M., Czerny, B., Ziolkowski, J., Gierliński, M., 2006, MNRAS, 370, 1534
\bibitem[\protect\citeauthoryear{Papadakis}{2004}]{Papadakis04}
Papadakis, I. E., 2004, MNRAS, 348, 207
\bibitem[\protect\citeauthoryear{Peterson}{2004}]{Peterson04}
Peterson, B. M.; Ferrarese, L.; Gilbert, K. M.; Kaspi, S.; Malkan, M. A.; Maoz, D.; Merritt, D. 2004, ApJ, 613, 682 
\bibitem[\protect\citeauthoryear{Pierce07}{2007}]{Pierce07}
Pierce, C. M.; Lotz, J. M.; Laird, E. S.; Lin, L.; Nandra, K.; Primack, J. R.; Faber, S. M.; Barmby, P.; Park, S. Q.; Willner, S. P.; and 9 coauthors 2007, ApJ, 660L, 19
\bibitem[\protect\citeauthoryear{Pizzella}{2005}]{Pizzella05}
Pizzella, A.; Corsini, E. M.; Dalla Bontà, E.; Sarzi, M.; Coccato, L.; Bertola, F.  2005, ApJ, 631, 785
\bibitem[\protect\citeauthoryear{Rahman}{2007}]{Rahman07}
Rahman, N., Howell, J. H., Helou, G., Mazzarella, J. M., Buckalew, B. 
2007, ApJ, 663, 895
\bibitem[\protect\citeauthoryear{Schombert}{1998}]{Schombert}
Schombert, J. 1998, AJ, 116, 1650
\bibitem[\protect\citeauthoryear{Sprayberry et al.}{1995}]{Sprayberry95}
Sprayberry, D., Impey, C. D., Bothun, G. D., Irwin, M. J., 1995, AJ, 109, 558
\bibitem[\protect\citeauthoryear{Turner et al.}{1999}]{Turner99}
Turner, T. J., George, I. M., Nandra, K.,  Turcan, D., 1999, ApJ, 524, 667
\bibitem[\protect\citeauthoryear{Uttley}{2002}]{Uttley02}
Uttley, P., McHardy, I. M., Papadakis, I. E. 2002, MNRAS, 332, 231
\bibitem[\protect\citeauthoryear{Vasudevan}{2009}]{Vasudevan2009}
Vasudevan, R.V., Fabian, A.C. 2009 MNRAS, 392 1124 
\bibitem[\protect\citeauthoryear{Vaughan et al.}{2003}]{Vaughan03}
Vaughan, S., Edelson, R., Warwick, R. S., Uttley, P. 2003, MNRAS, 345, 1271
\bibitem[\protect\citeauthoryear{Vaughan et al.}{2005}]{Vaughan05}
Vaughan, S., Iwasawa, K., Fabian, A. C., Hayashida, K. 2005, MNRAS, 356, 524
\bibitem[\protect\citeauthoryear{Ziolkowski}{2005}]{Ziolkowski05}
Ziolkowski, J., 2005, MNRAS, 358, 851
\end{thebibliography}
\end{document}